\documentclass[usenatbib]{mn2e}
\pdfoutput=1
\usepackage{graphicx}
\usepackage{amsmath}
\usepackage{placeins}
\usepackage{mathptmx}
\usepackage{txfonts}
\usepackage[T1]{fontenc}
\usepackage{ae,aecompl}
\usepackage{placeins}
\usepackage{hyperref}
\usepackage{pdfpages}
\hypersetup{colorlinks,
citecolor=blue
}

\title[Dynamical modelling of `UGC 7321' in braneworld gravity]
{Dynamical modelling of disc vertical structure in superthin galaxy `UGC 7321' in braneworld gravity: An MCMC study}

\author[Aditya, Banerjee, Banerjee \& Sengupta]
       {K. Aditya $^{1}$,\thanks{E-mail :kaditya@students.iisertirupati.ac.in} 
        Indrani Banerjee$^{2}$\thanks{E-mail :tpib@iacs.res.in},
       Arunima Banerjee$^1$\thanks{E-mail : arunima@iisertirupati.ac.in}
       and Soumitra Sengupta$^{2}$\thanks{E-mail :tpssg@iacs.res.in}  \\
$^1$  Department of Physics, Indian Institute of Science Education and Research (IISER) Tirupati, Tirupati - 517507, India \\
$^2$  School of Physical Sciences, Indian Association for the Cultivation of Science,2A \& 2B Raja S. C. Mullick Road, Kolkata-700032, India
}

\begin{document}
\maketitle

\begin{abstract}

Low surface brightness (LSBs) superthins constitute classic examples of very late-type galaxies, with their disc dynamics strongly regulated by their dark matter halos. In this work we consider a gravitational origin of dark matter in the brane world scenario, where the higher dimensional Weyl stress term projected onto the 3-brane acts as the source of dark matter. In the context of the braneworld model, this dark matter is referred to as the \emph{`dark mass'}.This model has been successful in reproducing the rotation curves of several low surface brightness and high surface brightness galaxies. Therefore it is interesting to study the prospect of this model in explaining the vertical structure of galaxies which has not been explored in the literature so far. Using our 2-component model of gravitationally-coupled stars and gas in the external force field of this \emph{dark mass}, we fit the observed scale heights of stellar and atomic hydrogen (HI) gas of superthin galaxy `UGC7321' using the Markov Chain Monte Carlo approach. We find that the observed scaleheights of `UGC7321' can be successfully modelled in the context of the braneworld scenario. In addition, the model predicted rotation curve also matches the observed one. The implications on the model parameters are discussed.

\end{abstract}

\section{Introduction}

Historically, the concept of dark matter was invoked to address the missing mass problem in spiral galaxies \citep{rubin1979extended} as well 
as to explain the mass discrepancy in galaxy clusters \citep{zwicky1937masses, zwicky1933rotverschiebung}. The observed rotation curves of 
galaxies, as determined by optical tracers or by HI 21cm radio-synthesis studies, are flat or tend to be asymptotically flat. Interestingly,
however, the observed distribution of visible matter predicts a Keplerian fall-off beyond the visible galactic disc \citep{binney2011galactic}. 
Basic physics suggests, the flatness of the rotation curve requires the total galactic mass to be increasing linearly with galacto-centric radius
even beyond the baryonic disc of the galaxy. This hinted at the presence of non-luminous matter in the discs of galaxies, the ``dark matter", 
which was invoked to explain the mass discrepancy and hence the flat rotation curves of spiral galaxies. In fact,  the ``dark matter" has been used to explain the observed rotation curves of a wide range of spiral galaxies including massive high surface brightness galaxies (HSBs), 
intermediate-mass low surface brightness galaxies (LSBs) to dwarf irregulars  \citep{sofue2001rotation,mcgaugh2001high, kranz2003dark, gentile2004cored, 
de2005halo, de2008high, oh2015high}. Similarly, the concept of ``dark matter" came as a rescue to address the mass discrepancy problem in galaxy clusters.
The mass of a galaxy cluster estimated by summing up the masses of its individual member galaxies is much lower than the virial mass of the galaxy-cluster 
determined using the observed line of sight velocity dispersion values of the member galaxies \citep{carlberg1997average}. This again hinted at the presence 
of non-luminous matter at cluster scales, which could be explained by invoking the concept of ``dark matter".

Although the ``dark matter" is generally
thought to resolve a host of astrophysical and cosmological problems,
the fundamental particles constituting the dark matter have evaded detection in dark matter search experiments 
(See, for example, Cryogenic Dark Matter Search (CDMS), \citet{agnese2013silicon}). 
The lack of detectional evidence 
for dark matter particles as well as a large number of problems arising in the
dark-matter particle approach \citep{Peebles:2010di}, \citep{kroupa2012dark,Kroupa:2014ria}, \citep{pawlowski2015persistence} opens up the possibility for a gravitational origin of dark matter wherein Newtonian gravity 
is modified to explain the `missing mass' problem. 

One of the earliest attempts to modify the Newton's laws for explaining 
the galactic rotation curves was Modified Newtonian dynamics (MOND) \citep{milgrom1983modification}, which is well-tested in 
the context of the Milky Way \citep{famaey2005modified} and also on a large sample of spiral 
galaxies \citep{de1998testing, sanders1998rotation, sanders2007confrontation}. 
Besides, extra dimensional models or brane-worlds can also lead to alternative gravitational origin 
of ``dark matter" \citep{Binetruy:1999ut, Csaki:1999jh, Mazumdar:2000gj, Maartens:2000fg, Maartens:2003tw, Koyama:2003be, Haghani:2012zq} 
where the Standard Model particles and fields are confined in a 3-brane while gravity enters into the bulk
\citep{Antoniadis:1990ew,Antoniadis:1998ig,ArkaniHamed:1998rs,Randall:1999vf,Garriga:1999yh,Randall:1999ee,Csaki:1999mp}, (see \cite{Fichet:2019owx} and references 
therein for braneworld models with matter fields not localized onto the brane). 
Higher dimensional models or brane-worlds were mainly introduced to provide a scheme to unify all the known forces of nature thereby
giving birth to string theory and eventually M-theory \citep{Kaluza:1921tu,Klein:1926fj,Horava:1995qa,Polchinski:1998rq}.
The huge difference between the Planck scale and the electroweak scale led to the gauge hierarchy problem in particle physics which
could also be addressed by introducing higher dimensions \citep{Antoniadis:1990ew,ArkaniHamed:1998rs,Antoniadis:1998ig,Randall:1999ee,Csaki:1999mp}. 
Further, extra-dimensional models have interesting phenomenological \citep{ArkaniHamed:1998nn, Davoudiasl:1999tf, Davoudiasl:1999jd, Davoudiasl:2000wi, Hundi:2011dc, 
Chakraborty:2014zya} and cosmological implications \citep{Dienes:1998hx,Lukas:1999yn, ArkaniHamed:1999gq,Mazumdar:2000sw,Chakraborty:2013ipa,
Banerjee:2017lxi,Banerjee:2018kcz}.
Moreover, since the nature of gravity in the high energy regime is unknown, it is often believed that the deviations from Einstein gravity may manifest
itself through the existence of extra dimensions.

The brane world model of \cite{maartens2004brane} considers a single 3-brane embedded in a five dimensional bulk. 
The modifications in the Einstein's equations arise mainly due to the non-local effects of the bulk Weyl tensor. A brane observer perceives 
this Weyl stress term like a fluid, known as the Weyl fluid, with its own energy density and pressure. It was shown by \cite{mak2004can}, 
\cite{harko2006galactic}, \cite{boehmer2007galactic}, \cite{rahaman2008galactic} and \cite{gergely2011galactic} that such a model successfully complies
with observed rotation curves of galaxies. The aim of this work is to explore the prospect of the brane world model in explaining the observed scale height
of stars and neutral hydrogen gas (HI) in galaxies. Towards this end, we consider the prototypical low surface brightness superthin galaxy, UGC7321, which was
found to be dark matter dominated at all radii \citep{banerjee2010dark}. In fact, \cite{banerjee2013some} have shown that the superthin vertical structure of the stellar 
disc crucially depends on its dense and compact dark matter halo. Therefore, in this work, we intend to constrain its dark matter density profile in the brane world scenario 
using the observed stellar and HI scale heights of UGC 7321. We derive the density profile of the Weyl fluid which arises in the brane world scenario to resemble the cored 
dark matter halo profile consistent with mass models of the low surface brightness galaxies \citep{de2005halo}. We use this derived density profile of the Weyl fluid to mimic 
the dark matter density profile in the 2-component model of gravitationally-coupled stars and gas in the external force field of the dark matter halo \citep{narayan2002vertical}. 
The observed stellar and HI scale heights are used to constrain the vertical stellar velocity dispersion and vertical HI velocity dispersion of the galactic disc in addition to 
the Weyl parameters. Finally, we check the consistency of the Weyl model with the observed rotation curve of UGC 7321.  

The paper is organised as follows: in \S 2, we describe the brane-world model and present the density profile derived from such a model
mimicking the characteristics of the dark matter. In \S 3, we present the 2-component model of the baryonic disc in the force field of the dark matter 
halo such that the Weyl fluid plays the role of the dark matter. We present our results in the context of the the LSB super thin galaxy UGC 7321 in \S 4 and 
conclude with a summary of our findings in \S 5. 

\section{Braneworld gravity: A possible proxy to dark matter}
\label{S2}
We consider a single 3-brane embedded in a 5-D spacetime (the bulk). We assume that the  Standard Model particles and fields are confined in the brane while gravity 
permeates the extra dimensions. The coordinates of the bulk are denoted by latin indices $x^a$, $a=0,1,...4$ while the brane coordinates are denoted by Greek indices $x^\mu$,
$\mu=0,1,2,3$.

{We consider Einstein gravity in the bulk described by the action,
\begin{equation}
\label{S21}
S_{bulk}=\int d^5x \sqrt{-\tilde{g}}  \bigg[\frac{R}{2\kappa_5^2} + \Lambda_5 + \delta(\phi)\mathcal{L}_{m}\bigg]
\end{equation}
where $\tilde{g}$ is the determinant of the bulk metric $\tilde{g}_{ab}$ , $R$ is the bulk Ricci scalar, $\kappa^{2}_{5} = 8\pi G_{5}$ is the five dimensional gravitational constant, $\Lambda_{5}$ is the bulk cosmological constant and
$\mathcal{L}_{m}$ is the matter Lagrangian in the brane representing standard model fields.

From the action (given by \ref{S21}) the gravitational field equations in the bulk are given by,
\begin{equation}
\label{S2-4}
R_{ij}-\frac{1}{2}G_{ij}R=\kappa^2_5 T_{ij}
\end{equation}
where $R_{ij}$ is the bulk Ricci tensor and $T_{ij}$ the energy-momentum in the bulk.
The bulk energy-momentum tensor can be written as,
\begin{align}\label{S2-5}
T_{ab} = - \Lambda_{5}G_{ab} + \delta(\phi) (- \lambda_{T}g_{\mu\nu} + \tau_{\mu\nu}) e_{a}^{\mu} e_{b}^{\nu}
\end{align}
where $\phi$ refers to the extra coordinate, the brane being located at $\phi=0$ and $e^{a}_{\mu}$ projects the quantities of the bulk onto the brane. The induced metric on the $\phi=0$ time-like hypersurface is denoted by $g_{\mu\nu}$. The negative vacuum energy density on the bulk $\Lambda_{5}$, the brane tension $\lambda_T$ and the brane energy-momentum tensor $\tau_{\mu\nu}$ are the sources of the gravitational field on the bulk. 

In this work we adopt the 3+1+1 covariant approach in brane worlds developed in \cite{Keresztes:2009hy,Keresztes:2009wb}. In this formalism, first the space-like extra coordinate $\phi$ is singled out and the 5-D spacetime is foliated by a family of 4-D time-like $\phi=\rm{constant}$ hypersurfaces, where the hypersurface designated by $\phi=0$ represents the brane. Next, in the 4-D time-like hypersurfaces the time coordinate is singled out and the 4-D spacetime is foliated in terms of the 3-D space-like $t=\rm{constant}$ hypersurfaces. The propagation along the extra-dimension and the temporal direction are associated with the integral curves $n^a=\frac{dx^a}{d\phi}$ and $u^a=\frac{dx^a}{d\tau}$ respectively, obeying the normalization conditions $1=n^an_a=-u^bu_b$ and the perpendicularity condition $n^c u_c=0$. The bulk metric $\tilde{g}_{ab}$ is connected to the brane metric $g_{ab}$ by the relation,
\begin{align}
\tilde{g}_{ab}=g_{ab}+n_a n_b
\label{1a}
\end{align}

It can be shown that the 5-D covariant derivative (denoted by $\tilde{\nabla}$) of $u^a$ and $n^a$ can be expressed in terms of several kinematic quantities, involving three scalars, four three-vectors and a symmetric, trace-free three-tensor, of which two scalars and one of the three vectors arise in common during the decomposition of both $\tilde{\nabla}_au_b$ and $\tilde{\nabla}_an_b$ \citep{Keresztes:2009wb}. Further, the 5-D Weyl tensor $\tilde{C}_{abcd}$ can be decomposed into several gravito-electro-magnetic quantities (involving scalars, vectors and tensors) which carry distinct geometrical meaning \citep{Keresztes:2009wb}.

Gauss-Codazzi equations which relate the Riemann tensor of the bulk to that of the brane by means of the projectors and the extrinsic curvature tensor are used to arrive at the effective four dimensional gravitational field equations.} The extrinsic curvature $K_{\mu\nu}$, which encodes the embedding of the brane into the bulk is associated with the covariant derivative of the normalized normals to the brane $n^a$. If the brane has an energy momentum tensor, the extrinsic curvature $K_{\mu \nu}$ exhibits a discontinuity across the brane.This discontinuity in the extrinsic curvature is related to the brane energy momentum tensor by means of Israel junction conditions and a symmetrical brane embedding ($Z_2$ symmetry) \citep{Keresztes:2009wb,Chakraborty:2015bja}

With the above considerations the effective four-dimensional gravitational field equations on the brane assume the form,
\begin{equation} 
\mathcal{R}_{\mu\nu} - \frac{1}{2}\mathcal{R}g_{\mu\nu} = -\Lambda_{4}g_{\mu\nu} + 8\pi G_{4}\tau_{\mu\nu} + \kappa^{4}_{5}\pi_{\mu\nu}- E_{\mu\nu}   \label{EE}
\end{equation} 
where
\begin{align}
\Lambda_{4} &= \frac{1}{2}\kappa^{2}_{5}\bigg{[}\Lambda_{5}+\frac{1}{6}\kappa^{2}_{5}\lambda^{2}_{T}\bigg{]} \label{5}\\
G_{4} &= \frac{\kappa^{4}_{5}\lambda_{T}}{48\pi} \label{6}\\
\pi_{\mu\nu} &=-\frac{1}{4}\tau_{\mu\alpha}\tau^{\alpha}_{\nu} + \frac{1}{12}\tau\tau_{\mu\nu} + \frac{1}{8}g_{\mu\nu}\tau_{\alpha\beta}\tau^{\alpha\beta} - \frac{1}{24}g_{\mu\nu}\tau^{2} \label{7}\\
E_{\mu\nu} &=\tilde{C}_{abcd}e^{a}_{\mu}n^{b}e^{c}_{\nu}n^{d} \label{8}
\end{align}

In \ref{EE}, $\Lambda_{4}$ and $G_{4}$ represent the 4-dimensional cosmological constant and gravitational constant respectively while $\mathcal{R}_{\mu\nu}$ and $\mathcal{R}$ refer to the Ricci tensor and Ricci scalar on the brane. The local effects of the bulk on the brane is encoded in the term $\pi_{\mu\nu}$ while $E_{\mu\nu}$, the electric part of the bulk Weyl tensor $\tilde{C}_{abcd}$ captures the non-local effect from the free bulk gravitational field. In \ref{5} the brane tension can be adjusted with the bulk cosmological constant to yield de-Sitter, anti de-Sitter or flat branes such that it serves as the fine balancing relation of the Randall-Sundrum single brane model \citep{Randall:1999vf, Shiromizu:1999wj}. The conservation of the matter energy-momentum tensor on the brane enables us to constrain $E_{\mu\nu}$  and $\pi_{\mu\nu}$ as $D_\nu E^\nu_\mu-\kappa_5^4 D_\nu \pi^\nu_\mu =0$, (where $D_{\nu}$ represents the brane covariant derivative).

The symmetry properties of the trace-free tensor $E_{\mu\nu}$ allow an irreducible decomposition of the tensor in terms of a given 4-velocity field $u^\mu$ \citep{Maartens:2001jx,Harko:2004ui},
\begin{align}
E_{\mu\nu} &= -k^{4}\bigg{[}U(r)(u_{\mu}u_{\nu} + \frac{1}{3}\zeta_{\mu\nu})  + 2Q_{(\mu}u_{\nu)}+ P_{\mu\nu}\bigg{]} \label{12}
\end{align}
 where $\zeta_{\mu\nu} = g_{\mu\nu} + u_{\mu}u_{\nu}$ is the induced metric associated with the 3-D space-like hypersurfaces orthogonal to $u^\mu$, $k = \frac{\kappa_{5}}{\kappa_{4}}$ and $\kappa^{2}_{4} = 8\pi G_{4}$. 
It is important to note that $\kappa_4^2=\kappa_5^4\lambda_T/6$ and we retrieve general relativity in the limit $\lambda_T^{-1}\rightarrow 0$ (\citep{Harko:2004ui}).
The scalar $U(r) = -\frac{1}{k^{4}}E_{\mu\nu}u^{\mu}u^{\nu}$ in \ref{12} is often known as the ``Dark Radiation" term, while $Q_\mu=\frac{1}{k^{4}}\zeta^{\alpha}_{\mu}E_{\alpha\beta}u^{\beta}$ represents a spatial vector and $P_{\mu\nu} = -\frac{1}{k^{4}}\big{[}\zeta^{\alpha}_{(\mu}\zeta^{\beta}_{\nu)} - \frac{1}{3}\zeta_{\mu\nu}\zeta^{\alpha\beta}\big{]}E_{\alpha\beta}$ consists of a spatial, tracefree, symmetric tensor.

Our goal in this work is to explain observations at the galactic scale, namely, the vertical scaleheight of stars and gas and the rotation curves of galaxies in the framework of braneworld gravity. We have already noted that in such a scenario, the bulk geometry induces an effective energy-momentum tensor ($E_{\mu\nu}$) on the brane which can play the crucial role of dark matter. The effect of baryons on the brane geometry is neglected and therefore we take $\tau_{\mu\nu}=\pi_{\mu\nu}=0$. 

As a result the gravitational field equations on the brane (given by \ref{EE}) reduce to,
\begin{align} 
\mathcal{R}_{\mu\nu} - \frac{1}{2}\mathcal{R}g_{\mu\nu} = -\Lambda_{4}g_{\mu\nu} - E_{\mu\nu}   \label{EF}
\end{align}
The conservation of energy-momentum tensor on the brane then assumes the form $D_\nu E^\nu_\mu=0$. Moreover, the mass distribution of the dark matter can be approximately taken to be spherically symmetric and time-independent, and as a result
we take $Q_\mu=0$ in \ref{12} such that $D_\nu E^\nu_\mu=0$ leads to,
\begin{align}
\frac{1}{3}\bar{D}_\mu U +\frac{4}{3}U A_\mu + \bar{D}^\nu P_{\nu\mu} + A^\nu P_{\nu\mu}=0 \label{13}
\end{align}
where $A_\mu=u^\nu D_\nu u_\mu$ is the acceleration and $\bar{D}$ denotes covariant derivative on the space-like hypersurface associated with the metric $\zeta_{\mu\nu}$ orthonormal to $u_\mu$.
Further, assumption of spherical symmetry, enables us to write $A_\mu=A(r)r_\mu$, while the term $P_{\mu\nu}$ takes the form,
\begin{align}
P_{\mu\nu}=P(r)\left(r_\mu r_\nu-\frac{1}{3}\zeta_{\mu\nu}\right) \label{14}
\end{align} 
where $A(r)$ and $P(r)$ (also known as the ``Dark Pressure") are scalar functions of the radial coordinate $r$ and $r_\mu$ is the unit radial vector. In what follows we will derive static and spherically symmetric solution of \ref{EF}.

\subsection{Motion of test particles in the braneworld model}
\label{S2-1}
In this work we are interested in exploring the properties of low surface brightness galaxies (LSBs) where the dark matter halo dominates the disc dynamics even in the inner galaxy where the baryonic disc is present. \citep{Gergely:2011df, de2001mass, banerjee2010dark, banerjee2017mass}. Therefore, a static and  spherically symmetric spacetime is used to describe the brane metric and one can neglect the effect of the cylindrically symmetric baryonic disk on the brane geometry.

In the last section, we noted that the presence of extra dimensions endows an effective energy-momentum tensor on the brane $E_{\mu\nu}$ due to the non-local effects of the 
bulk Weyl tensor. This Weyl stress term essentially plays the role of dark matter in this work. Since LSBs are dark matter dominated, one can neglect the effect of the 
cylindrically symmetric baryonic disk on the brane geometry such that a static and spherically symmetric spacetime is used to describe the brane metric,

\begin{align}
ds^2=-e^{\nu(r)}dt^2 + e^{\lambda(r)}dr^2 + r^2(d\theta^2 +\rm{sin}^2 \theta d\phi^2 )\label{15}
\end{align}
We solve for $\nu(r)$, $\lambda(r)$, $U(r)$ and $P(r)$ from the fact that \ref{15} satisfies \ref{EF} and \ref{13}.

By studying the motion of test particles in the above spacetime it can be shown that the tangential velocity or the circular velocity of motion is given by 

\begin{align}
v_{c}^2=\frac{r\nu^\prime}{2} \label{16}
\end{align}
\citep{gergely2011galactic}

With \ref{15} the gravitational field equations and the energy momentum tensor conservation in the brane gives us, 
\begin{align}
&-e^{-\lambda}\bigg(\frac{1}{r^2}-\frac{\lambda^\prime}{r}\bigg)+\frac{1}{r^2}=3\alpha U   \\
&e^{-\lambda}\bigg(\frac{\nu^\prime}{r}+\frac{1}{r^2}\bigg)-\frac{1}{r^2}=\alpha (U+2P)    \\
& \frac{e^{-\lambda}}{2}\bigg(\nu^{\prime \prime}+  \frac{\nu^{\prime 2}}{2}+\frac{\nu^{\prime}-\lambda{^\prime}}{r}-\frac{\nu^{\prime}\lambda^{\prime}}{2}\bigg)                  =\alpha (U-P)\\
&\nu^{\prime} = - \frac{U^{\prime} + 2P^{\prime}}{2U + P} - \frac{6P}{(2U + P)r}
\end{align}
where prime denotes derivative with respect to $r$ and $\alpha = \frac{1}{4\pi G_{4}\lambda_{T}}$.
One can show that the solution of these equations lead to the following form for $e^{-\lambda}$,
\begin{align}
e^{-\lambda} &= 1 - \frac{\Lambda_{4}}{3}r^{2} - \frac{Q(r)}{r} - \frac{C}{r} \label{16-1}
\end{align}
where ${C=2GM}$ (${M}$ being the baryonic mass) and $Q(r)$ is defined as,
\begin{align}
Q(r) = \frac{3}{4\pi G_{4}\lambda_{T}}\int r^{2}U(r)dr \label{16-2}
\end{align}
 \citep{gergely2011galactic}.
From the form of $e^{-\lambda}$ it can be inferred that $Q(r)$ is the gravitational mass originating from the dark radiation and can be interpreted as the ``dark mass" term.

Further, one can show that for a static, spherically symmetric spacetime the ordinary differential equations for dark radiation $U(r)$ and dark pressure $P(r)$ satisfy ,
\begin{align}
\frac{dU}{dr} = -2\frac{dP}{dr} - 6\frac{P}{r} - \frac{(2U + P)[2G_{4}M + Q + \{ \alpha(U + 2P+\frac{2}{3} \chi) \}r^{3}]}{r^{2}\big{(} 1 - \frac{2G_{4}M}{r} - \frac{Q(r)}{r} - \frac{\chi }{3}r^{2}\big{)}} \label{17}
\end{align}
and
\begin{align}
\frac{dQ}{dr} = 3\alpha r^{2}U. \label{18}
\end{align}
where $\alpha = \frac{1}{4\pi G_{4}\lambda_{T}}$ and $\chi = -\Lambda_{4}$ \citep{gergely2011galactic}. 
In the subsequent calculations we will neglect the effect of the cosmological constant $\Lambda_{4}$ \citep{gergely2011galactic} on the vertical scale height of the galaxies, i.e. we will take $\chi = -\Lambda_{4}=0$. Since the observed cosmological constant required to explain the accelerated expansion of the universe is extremely small ($\Lambda_{4}\approx 10^{-52} \rm m^{-2}~ or~ 10^{-122} $ in Planckian units), its effect on the the mass energy of the galaxy can be ignored as being several orders of magnitude smaller than the observed masses.  

\ref{17} and \ref{18} can be recast into a more convenient form namely,
\begin{align}
\frac{d\mu}{d\theta} &= -(2\mu + p)\frac{\tilde{q} + \frac{1}{3}(\mu + 2p) }{1 - \tilde{q} } - 2\frac{dp}{d\theta} + 2\mu - 2p \label{19}\\
&=-2v^2_{tg}(2\mu+p)- 2\frac{dp}{d\theta} + 2\mu - 2p \nonumber \\
\frac{d\tilde{q}}{d\theta} &= \mu - \tilde{q} \label{20} 
\end{align} 
by defining the variables, 
\begin{align}
\tilde{q} = \frac{2G_{4}M + Q}{r} ;~~~\mu = 3\alpha r^{2}U ; ~~~ p= 3\alpha r^{2}P ; ~ ~~\theta = ln ~ r ; \label{20-1}
\end{align}
\ref{19} and \ref{20} can be referred to as the differential equations governing the source terms on the brane, while the circular velocity of the test particle $v_c$ assumes the form,
\begin{align}
 v^2_{c}=\frac{1}{2}\frac{\tilde{q}+\frac{1}{3}(\mu+2p)}{1-\tilde{q}} \label{20-2}
\end{align}

\subsection{Choice of the equation of state of the Weyl fluid}
\label{S2-2}
The brane observer perceives the extra dimensions through the term $E_{\mu\nu}$ which in turn can be written in terms of the dark radiation $U(r)$ and the dark pressure $P(r)$. These are like the energy density and the pressure of the stress energy tensor of the Weyl fluid whose origin is attributed to extra dimensions. An equation of state connecting the dark radiation $U$ and dark pressure $P$ is therefore necessary since the source equations \ref{19} and \ref{20} cannot be solved simultaneously until we impose some further conditions on them. 
Hence, we choose some specific relations between dark radiation $U$ and dark pressure $P$, necessarily defining the various equations of state in the framework of the brane world model. 

In order to close the system of field equations \ref{19} and \ref{20}, we adopt the 3+1+1 covariant approach developed in \cite{Keresztes:2009hy,Keresztes:2009wb}, also discussed briefly in \ref{S2}. This approach is quite different from the full 5-D approach used in \cite{Wiseman:2001xt} where the five dimensional Einstein's equations were numerically solved assuming that the matter localized in the brane is associated with a static and spherically symmetric configuration. Such a choice of matter distribution on the brane was considered since \cite{Wiseman:2001xt} aimed at explaining observations related to stars confined in the brane. In order to obtain a regular bulk metric \cite{Wiseman:2001xt} assumed a density profile resembling a deformed top-hat function which resulted in a regular bulk geometry with axial symmetry. Our aim on the other hand, is to explain observations at the galactic scale where the bulk geometry plays the vital role of dark matter through the term $E_{\mu\nu}$.

In order to proceed further we consider the following assumptions: (i) no matter in the brane (the effect of baryons on the brane metric is neglected), (ii) fine-tuning on the brane (i.e. $\Lambda_4=0$), (iii) cosmological vacuum in the 5-D spacetime, (iv) symmetrical embedding on the brane ($Z_2$ symmetry) and (v) the brane spacetime is static and spherically symmetric, such that the brane metric assumes the form \ref{15}.

From the above assumptions it can be shown that the only non-zero kinematical quantity associated with the temporal normal $u^\mu$ is the acceleration $A_\mu=u^\nu D_\nu u_\mu$ \citep{Gergely:2011df}. Since we are interested in static and spherically symmetric solutions $Q_\mu$ in \ref{12} vanishes and the requirement of spherical symmetry ensures that the brane spacetime can be further decomposed into a 2+1+1 form such that $A_\mu=A(r)r_\mu$ and $P_{\mu\nu}$ in \ref{12} can be written as \ref{14},
where $A(r)$ and $P(r)$ (also known as the ``Dark Pressure") are scalar functions of the radial coordinate $r$.
Due to spherical symmetry the electric part of the 4-D Weyl tensor $\tilde{E}_{\mu\nu}=C_{\alpha\mu\beta\nu}u^\alpha u^\beta$ ($C_{\alpha\mu\beta\nu}$ being the 4-D Weyl tensor) can be further reduced in the form,
\begin{align}
\tilde{E}_{\mu\nu}=\tilde{E}(r)\bigg(r_\mu r_\nu-\frac{\zeta_{\mu\nu}}{3}\bigg) 
\label{15a}
\end{align}   
Moreover, we have the kinematical scalar $\tilde{\Theta}(r)=\bar{D}_ar^a$ associated with the expansion of the radial geodesics in the $t=\rm constant$ hypersurfaces.

The five variables: $U(r)$, $P(r)$, $\tilde{E}(r)$, $A(r)$ and $\tilde{\Theta}(r)$, discussed above are associated with four independent field equations \citep{Gergely:2011df,Keresztes:2009wb}, namely,
\begin{align}
k_4^4 U=\tilde{\Theta}\bigg(A-\frac{\tilde{\Theta}}{4}\bigg)+\frac{4\tilde{E}}{3} +\frac{1}{r^2}
\label{16a}
\end{align}
\begin{align}
k_4^4 P=\tilde{\Theta}\bigg(A+\frac{\tilde{\Theta}}{2}\bigg)-\frac{2\tilde{E}}{3} -\frac{2}{r^2}
\label{17a}
\end{align}
\begin{align}
\frac{r\tilde{\Theta}}{2}\tilde{\Theta}^\prime + \frac{\tilde{\Theta}^2}{2}+\frac{4\tilde{E}}{3} + A\tilde{\Theta}=0
\label{18a}
\end{align}
\begin{align}
\frac{r\tilde{\Theta}}{2}A^\prime + A^2 + \frac{\tilde{\Theta}^2}{4} -\frac{4\tilde{E}}{3}-\frac{1}{r^2}=0
\label{19a} 
\end{align}
where prime denotes derivative with respect to the radial coordinate.
It is important to note that \ref{18a} and \ref{19a} involve only variables associated with the brane dynamics. 
It turns out that in the Schwarzschild scenario, the variables $\tilde{E}$, $\tilde{\Theta}$ and $A$ can be related in the following way \citep{Gergely:2011df},
\begin{align}
\frac{2\tilde{E}}{3}+A\tilde{\Theta}= \tilde{\Theta}\bigg(\frac{\tilde{\Theta}}{4}+A\bigg)-\frac{1}{r^2}=0
\label{20a}
\end{align}
\ref{20a} ensures that \ref{18a} and \ref{19a} coincide and the variables $\tilde{E}$, $\tilde{\Theta}$ and $A$ are still determined by three equations \cite{Gergely:2011df}.
In the presence of the Weyl fluid, the resultant spherically symmetric metric exhibits a departure from the Schwarzschild spacetime and consequently a modification in \ref{20a} is considered,
\begin{align}
\frac{2\tilde{E}}{3}+A\tilde{\Theta}= M_0\tilde{\Theta}\bigg(\frac{\tilde{\Theta}}{4}+A\bigg)-\frac{N_0}{r^2}
\label{21a}
\end{align}
where $M_0$ and $N_0$ are two constant deformation parameters characterizing the Weyl fluid and reducing to 1 in the event the spherically symmetric brane is described by the Schwarzschild solution. \ref{18a}, \ref{19a} and \ref{21a} comprises of three equations in the three variables $A(r)$, $\tilde{\Theta}(r)$ and $\tilde{E}(r)$ and the Cauchy-Peano theorem ensures the existence of a solution.

Substitution of \ref{16a} and \ref{17a} in \ref{21a} gives rise to an equation of state connecting the ``dark radiation" and the ``dark pressure" on the  brane,
\begin{align}
P(r)=(a-2)U(r)-\frac{B}{k_4^4 r^2} 
\label{22a}
\end{align}
where, $M_0=a/(2a-3)$ and $N_0=(a-B)/(2a-3)$ such that in the Schwarzschild limit $a=3$ and $B=0$.
In terms of the reduced variables defined in \ref{S2-1}, the above equation of state can be rewritten as, 
\begin{align}
p(\mu)=(a-2)\mu-B 
\label{23a}
\end{align}
This equation also serves as a closure condition, which enables us to solve \ref{19} and \ref{20} consistently. The structure of the resultant metric in the brane is discussed in the next section.

Having outlined the procedure by which the 3+1+1 covariant dynamics can be used as a theoretical framework to solve \ref{19} and \ref{20}, we now address the consistency of the solution obtained in the brane with that of the bulk. 
In order to understand the evolution of the various kinematic and gravito-electro-magnetic quantities (mentioned earlier and also discussed in \cite{Keresztes:2009wb}) along the temporal and extra-dimension, one needs to consider the Ricci identities discussed in \cite{Keresztes:2009wb} which are first order differential equations describing the evolution of the aforesaid quantities along the integral curves associated with $n^a$ and $u^a$. Since we are considering a static scenario, \ref{23a} which serves as a boundary condition to be satisfied on the brane at some arbitrary time, continues to hold for all times.
On the other hand while considering off-brane evolution, the condition \ref{23a} serves as an initial condition to be satisfied on the brane and necessary to solve the first order differential equations describing the evolution along the bulk. The existence of the solution of such differential equations is again guaranteed due to the Cauchy-Peano theorem. 

One however needs to ensure that the condition \ref{23a} is consistent with the constraint equation on the brane, i.e. $D_\mu E^\mu_\nu=0$ (given by \ref{13}). For a static and spherically symmetric brane \ref{13} simplifies to,
\begin{align}
\frac{r\tilde{\Theta}}{2}(U+2P)^\prime + 4 A U + P(2A + 3 \tilde{\Theta})=0
\label{24a}
\end{align}
which can be shown to follow from \ref{16a}-\ref{19a}. Therefore, a static and spherically symmetric background ensures that \ref{24a} will always hold true on the brane and hence the intial condition given by \ref{23a} is allowed. Hence, the equation of state chosen in this work to close the system of equations \ref{19} and \ref{20}, is consistent with the evolution along the bulk. The additional advantage of choosing \ref{23a} is that it assumes a simple linear relation connecting the ``dark radiation" and the ``dark pressure" and can successfully explain the flat rotation curves observed in the galaxies. It turns out that such a choice of the equation of state can also explain the vertical scaleheight data of the LSB galaxies, which we show in the subsequent sections.

\subsection{Density profile for the Weyl fluid}
\label{S2-3}
Using the equation of state discussed in the last section and ignoring the effect of the cosmological constant, the equation for the dark pressure \ref{19} can be simplified to ,
\begin{align}
(2a-3)\frac{d\mu}{d\theta}=-\frac{(a\mu-B)(\tilde{q}+(2a-3)\mu/3-2B/3)}{1-\tilde{q}}+2\mu(3-a)+2B \label{22}
\end{align}
while the reduced dark radiation assumes the form,
\begin{align}
&\mu(\theta)=\theta^{2(3-a)/(2a-3)}\rm{exp}\bigg[-\frac{2a}{2a-3}\int v^2_{c}(\theta)d\theta   \bigg]\times \nonumber \\
&\bigg \lbrace C -\frac{3B}{2a-3}\int [1+v_c^2(\theta)]\theta^{-2(3-a)/(2a-3)}\times \nonumber \\
& \rm{exp}\bigg[\frac{2a}{2a-3}\int v_c^2 (\theta) d\theta\bigg]   \bigg \rbrace \label{23}
\end{align}
where $C$ is an arbitrary integration constant \citep{gergely2011galactic}.

In what follows, we will consider the situation where $a\ne 3/2$ and $\tilde{q}<<1$. The justification for $\tilde{q}<<1$ arises from the fact that a typical galactic dark matter halo has mass $ M  \approx 10^{12}M_\odot$ and radius $R  \approx 100$ kpc, such that $ {\tilde{q}\approx \frac{Q(r)}{R}\approx 10^{-7}<< 1}$. Since observations reveal that in a galaxy mass is directly proportional to the radius, this ratio remains roughly constant for all galactic radii. The smallness of $\tilde{q}$ further enables us to neglect higher order terms in $\tilde{q}$, such that \ref{20} assumes the form,

\begin{align}
\frac{d^2\tilde{q}}{d\theta^2}+m\frac{d\tilde{q}}{d\theta}-n\tilde{q}=b  
\label{24}
\end{align}

\begin{align} 
 m=1-\frac{B}{3}-\frac{2}{3}\frac{a(B-3)+9}{2a-3}~~~~a\neq \frac{3}{2}
\label{25}
\end{align}

\begin{align} 
n=\frac{2}{3}\frac{a(2B-3)+9}{2a-3} ~~~~a\neq \frac{3}{2}
\label{26}
\end{align}

\begin{align}  
 b=\frac{2}{3}\frac{B(B-3)}{3-2a}   ~~~~a\neq \frac{3}{2} 
\label{27}
\end{align}
The general solution of \ref{24} is,
\begin{align} 
\tilde{q}(r)=q_0+ C_1 r^{l_1} + C_2 r^{l_2}
\label{28}
\end{align}
where $C_1$ and $C_2$ are constants of integration and $q_0$ is given by,
\begin{align} 
q_0=-\frac{b}{n}=\frac{B(B-3)}{a(2B-3)+9}
\label{29}
\end{align}
while 
\begin{align}
 l_{1,2}=\frac{-m\pm\sqrt{m^2+4n}}{2}
\label{30}
\end{align}
The solution for reduced dark radiation is given by,
\begin{align}
\mu(r)=q_0+C_1(1+l_1)r^{l_1}+C_2(1+l_2)r^{l_2} 
\label{31}
\end{align}

In the original radial coordinate $r$ the solution for dark radiation $U(r)$ is,
\begin{align}
 \rho_h(r)=3\alpha U(r)=\frac{q_0}{r^2}+C_1(1+l_1)r^{l_1-2}+C_2(1+l_2)r^{l_2-2} \label{32}
\end{align}
which serves as the proxy for the density profile of dark matter.

The dark mass profile is given by,
\begin{align}
Q(r)=r(q_0+C_1r^{l_1}+C_2r^{l_2})-2GM_0
\label{33}
\end{align}
 where $M_0$ is a mass like constant due to the Weyl fluid.

For completeness we also mention that the tangential velocity of a test particle in the `dark matter' dominated region is given by ,
\begin{align}
v_c^2\approx v_{c_\infty}^2+\gamma r^{l_1} +\eta r^{l_2} ~~~\rm{where,}
\label{34}
\end{align}

\begin{align}
v_{c_\infty}^2=\frac{1}{3} (a q_0-B) 
\label{35}
\end{align}

\begin{align}
\gamma=\frac{C_1}{2}\bigg[1+\frac{(2a-3)}{3}(1+l_1)   \bigg]
\label{36}
\end{align}

\begin{align}
\eta=\frac{C_2}{2}\bigg[1+\frac{(2a-3)}{3}(1+l_2)\bigg]
\label{37}
\end{align}
\citep{Gergely:2011df}. At this stage, it is important to mention the components of the metric given in \ref{15}. The $g_{tt}$ component of the metric can be obtained by solving \ref{16} where the tangential velocity $v_c$ of the test particle is given by \ref{34} which yields,
\begin{align}
{e^{\nu(r)}=C_\nu r^{2 v_{c_\infty}^2}\times exp \bigg[C_1 \frac{3+(2a-3)(1+l_1)}{3l_1}r^{l_1} +
 C_2\frac{3+(2a-3)(1+l_2)}{3l_2}r^{l_2}}\bigg] \label{37a}
\end{align}
where $C_\nu$ is an arbitrary constant of integration. In the outer parts of the galaxy with large radial distances one can approximate $e^{\nu(r)} \approx C_\nu r^{2v_{c_\infty}^2}$ (\cite{Gergely:2011df}). The $g_{rr}$ component of the metric is given by \ref{16-1}, where $Q(r)$ can be obtained from \ref{33} and contribution of $\Lambda_4$ is neglected (discussion in \ref{S2-1}) such that,
\begin{align}
{e^{-\lambda} = 1 - (q_0+C_1r^{l_1}+C_2r^{l_2})+\frac{2GM_0}{r} - \frac{C}{r}} \label{37-b}
\end{align}

In order to obtain a flat rotation curve at large distances $l_1$ and $l_2$ should be negative. The constrain on the Weyl parameters from the rotation curve has been derived in \cite{gergely2011galactic}. The goal of this work is to constrain the Weyl parameters from the vertical scale height data of the Low Surface Brightness galaxies (LSBs).

One can show that when $\tilde{q}<<1$ and $a\ne3/2$ the parameters $m$, $n$, $q_0$ and $v^2_{c_\infty}$ can be further simplified such that,

\begin{align}
m\approx \frac{4a-9}{2a-3},
\label{38}
\end{align}

\begin{align}
n\approx -2\frac{a-3}{2a-3},
\label{39}
\end{align}
\begin{align}
q_0\approx \frac{B}{a-3}~~~\rm{and}
\label{40}
\end{align}
\begin{align}
v^2_{c_\infty}\approx \frac{a}{3} \bigg(q_0-\frac{B}{a}	\bigg)
\label{41}
\end{align}
\ref{38} and \ref{39} implies that
\begin{align}
l_1\approx -1 ~~~\mathrm{and} ~~~l_2\approx -1+\frac{3}{2a-3}
\label{42}
\end{align}
which further ensures that $a$ cannot assume values between $3/2$ to $3$. The positivity of $v^2_{c_\infty}$ requires that when $a<3/2$, $B\leq 0$ while when $a>3$, $B>0$. 
The density profile for the Weyl fliud assumes the form,
\begin{align}
\rho_h(r)\approx \frac{q_0}{r^2}+\frac{3C_2}{2a-3}r^{-3(1-\frac{1}{2a-3})}
\label{43}
\end{align} 
while the rotation curve is given by,
\begin{align}
v_c^2\approx \frac{B}{a-3}+ \frac{C_1}{2}r^{-1}+C_2 r^{-1+\frac{3}{2a-3}}
\label{44}
\end{align}
By taking $C_1=\frac{2G(M_b+M_U)}{c^2}$ {(where ${M_U=Q(r)/G}$ and ${M_b}$ here refers to the baryonic mass)}, $C_2=C c^2 R_{c(DM)}^{1-\alpha_{DM}}=-\beta_{DM} $ and by defining $\alpha_{DM}=3/(2a-3)$ and $\beta_{DM}=B/(a-3)$, the final expressions for the rotation curve and the density profile are given by,

\begin{align}
\bigg(\frac{v_c(r)}{c}\bigg)^2\approx \frac{G(M_b+M_U)}{c^2 r}+\beta_{DM}\bigg[1 -\bigg(\frac{R_{c(DM)}}{r}\bigg)^{1-\alpha_{DM}}\bigg]~~~\rm{and}
\label{45}
\end{align}

\begin{align}
\rho_h(r)\approx \frac{c^2\beta_{DM}}{Gr^2}\bigg[1-\alpha_{DM}\bigg( \frac{R_{c(DM)}}{r}\bigg)^{1-\alpha_{DM}}\bigg] 
\label{46}
\end{align} 
Due to the constraints on the parameters $a$ and $B$, it can be shown that either $\alpha_{DM}<0$ or $0<\alpha_{DM}<1$ and $0<\beta_{DM}<<1$ \citep{gergely2011galactic}.

It is interesting to compare the Weyl model derived from braneworld gravity with other alternatives to dark matter, e.g. MOND (Modified Newtonian dynamics)\citep{milgrom1983modification}. Within the domain of MOND, there exists a universally constant acceleration scale below which the gravitational dynamics deviates significantly from Newton's laws. The framework of MOND becomes specially relevant in the low acceleration regimes e.g. stars in the outer parts of galaxies such that the model can successfully reproduce the flat rotation curves observed in galaxies without invoking dark matter. The modifications introduced in the laws of gravitation lead to a relation between the asymptotic velocity of galaxies and the total baryonic mass. Such a relation commonly known as the Baryonic Tully Fisher relation entails that the fourth power of the asymptotic velocity of galaxies is proportional to the total baryonic mass.

It may be interesting to investigate the consistency of the rotation curve derived in the Weyl model \ref{45} with the Baryonic Tully Fisher Relation (BTFR) and the Radial Acceleration Relation (RAR) \citep{McGaugh:2000sr,McGaugh:2011ac,McGaugh:2016leg}.
The Baryonic Tully Fisher Relation (BTFR) states that the fourth power of the asymptotic velocity of galaxies $V_f$ is proportional to the total baryonic mass $M_B$, i.e.,
\begin{align}
M_B=A V_f^4 \label{Eq1}
\end{align}
where $V_f$ is obtained from the flat part of the galactic rotation curve. By investigating the mass distribution and the rotation curves of a large number of galaxies with different morphologies, the proportionality constant $A$ was shown to vary between $35-50 \rm ~ M_\odot km^{-4}s^4$ \citep{McGaugh:2000sr,McGaugh:2011ac}.    
This relation can be derived in the deep MOND regime \citep{milgrom1983modification}, where the gravitational force $F$ is assumed to be proportional to the square of the acceleration $a$, 
\begin{align}
{F=\frac{GM_{tot}}{r^2}=\frac{a^2}{g^\dagger}=\frac{1}{g^\dagger}\bigg(\frac{v^2}{r}\bigg)^2}  \label{Eq2}
\end{align}
where $v$ is the circular velocity at radius $r$ and $g^\dagger$ is considered to be a universally constant acceleration scale such that $a<<g^\dagger$ defines the deep MOND regime. Thus, if the force law given by \ref{Eq2} is assumed, it can be easily shown that \ref{Eq1} follows from \ref{Eq2}. 
The Radial Acceleration Relation (RAR) proposes that the observed radial acceleration $g_{_{obs}}$ derived from the rotation curve and the baryonic acceleration $g_{_{bar}}$ derived from the observed mass distribution of galaxies obeys the following relation \cite{McGaugh:2016leg},
\begin{align}
g_{_{obs}}=\frac{g_{_{bar}}}{1-e^{-\sqrt{g_{_{bar}}/g^\dagger}}}
\label{Eq3}
\end{align}
It can be shown from \ref{Eq3} that when $g_{_{bar}}<<g^\dagger$, $g_{_{obs}}\propto \sqrt{g_{_{bar}}}$ while $g_{_{obs}}\approx g_{_{bar}}$ at high accelerations \cite{McGaugh:2016leg}. 
By analysing data of 153 galaxies from the SPARC database, it was reported that $g^\dagger =\rm 1.2 \pm 0.02 ~(random) \pm 0.24~ (systematic) \times 10^{-10} \rm m~ s^{-2} $ \citep{McGaugh:2016leg,Lelli:2017vgz}. 

In what follows we will show that if we assume the BTFR then we can obtain the RAR from the circular velocity profile derived in the Weyl model.  
It is clear from \ref{45} that at large $r$, the rotation curve is dominated by the term $\beta_{DM}$, i.e. $v_c\approx c \sqrt{\beta_{DM}} $ at the outer part of the galaxy. This parameter which is inherited from the Weyl model is therefore fixed from the flat part of the observed rotation curve of the galaxies, which implies,
\begin{align}
v_c^2=V_f^2=c^2 \beta_{DM}  \label{Eq4}
\end{align} 
At this point if we assume that the galaxy obeys the BTFR, then from \ref{Eq1} and \ref{Eq4},
\begin{align}
V_f^4=\frac{M_B}{A}=c^4\beta_{DM}^2 \label{Eq5}
\end{align}
This gives a connection between the Weyl parameter $\beta_{DM}$ with the total baryonic mass of the galaxy, such that,
\begin{align}
M_B=A c^4\beta_{DM}^2 \label{Eq6}
\end{align}
Now, in order to derive the RAR, we rewrite \ref{45} in the following form,
\begin{align}
\frac{v_c^2}{r}\approx \frac{GM_b}{r^2} + \frac{GM_U}{r^2} + c^2 \frac{\beta_{DM}}{r}\bigg[1 -\bigg(\frac{R_{c(DM)}}{r}\bigg)^{1-\alpha_{DM}}\bigg]  \label{Eq7}
\end{align}  
The left hand side of \ref{Eq7} can be identified with the observed radial acceleration at each radius ($g_{_{obs}}$) derived from the rotation curve, while the radial acceleration solely due to the baryons is given by $g_{_{bar}}=\frac{GM_b}{r^2}$. This enables us to rewrite \ref{Eq7} in the folowing form,
\begin{align}
\label{Eq8}
g_{_{obs}}&=g_{_{bar}} +\frac{G M_U}{r^2}+c^2 \frac{\beta_{DM}}{r}\bigg[1-\bigg(\frac{R_{c(DM)}}{r}\bigg)^{1-\alpha_{DM}}\bigg]\nonumber \\
&=g_{_{bar}} +\frac{g_{_{bar}}M_U}{M_b}+\beta_{DM}c^2\sqrt{\frac{g_{_{bar}}}{GM_b}}\bigg[1-\bigg( \sqrt{\frac{g_{_{bar}}}{GM_b}} R_{c(DM)}\bigg)^{1-\alpha_{DM}}\bigg] \nonumber \\
\end{align}
\ref{Eq8} can be obtained from \ref{Eq7} by replacing $1/r$ in terms of $g_{_{bar}}$ and $M_b$. If we assume that $M_U<< M_b$ (i.e. the core is dominated by the baryons), the second term in \ref{Eq8} can be dropped compared to the other terms. 

Further, at large $r$, $M_b$ becomes equal to the total baryonic mass $M_B$ such that \ref{Eq8} can be written as,
\begin{align}
g_{_{obs}}&=g_{_{bar}} + \sqrt{\frac{g_{_{bar}}}{GA}}\bigg[1-\bigg( \sqrt{\frac{g_{_{bar}}}{GA}}\frac{R_{c(DM)}}{c^2\beta_{DM}}\bigg)^{1-\alpha_{DM}}\bigg]
\label{8a}
\end{align}

It is important to note that in \ref{8a} the quantity $\frac{GAc^4\beta_{_{DM}}^2}{R^2_{c(DM)}}$ defines an acceleration scale. Recalling that $\beta_{DM}<<1$ and considering representative values of $\beta_{DM}\approx 10^{-7}$, $R_{c(DM)}\approx 1.5 \rm ~ kpc$ and $A\approx 35 ~\rm M_\odot km^{-4}s^4$, it can be shown that $\frac{GAc^4\beta_{_{DM}}^2}{R^2_{c(DM)}}\approx 1.76\times 10^{-10} \rm m~ s^{-2}$ which is very close to the universal acceleration scale $g^\dagger$ of MOND, where the estimated value of $g^\dagger =\rm 1.2 \pm 0.02 ~(random) \pm 0.24~ (systematic) \times 10^{-10} \rm m~ s^{-2} $ \cite{McGaugh:2016leg}.
We further note that $\alpha_{DM}$ has the theoretical constraints, either $\alpha_{DM}<0$ or $0<\alpha_{DM}<1$. Therefore, in the limit, $g_{_{bar}}<<\frac{GAc^4\beta_{_{DM}}^2}{R^2_{c(DM)}}$ and taking $\alpha_{DM}<0$, the second term in the square bracket of \ref{8a} can be easily dropped compared to unity. Hence, in such a situation \ref{8a} reduces to,
\begin{align}
g_{_{obs}}\approx g_{_{bar}} + \sqrt{\frac{g_{_{bar}}}{GA}}
\label{Eq9}
\end{align}
Moreover, considering $A\approx 35 ~\rm M_\odot km^{-4}s^4$ \cite{McGaugh:2000sr}, it turns out that $GA\approx 4.67\times 10^9 \rm ~m^{-1}s^2$.
Since we are in the regime $g_{_{bar}}<< 10^{-10} \rm m~ s^{-2}$, the second term in \ref{Eq9} dominates over the first and we recover the result $g_{_{obs}} \propto g_{_{bar}}^{1/2} $ as expected in the deep MOND regime.

In the event $r\approx R_{c(DM)}$ or $\alpha_{DM} \to 1$, the term in the square bracket of \ref{Eq8} becomes negligible compared to the first term such that the Newtonian relation $g_{_{obs}}=g_{_{bar}}$ is recovered.
This discussion explains that if BTFR is assumed then the RAR can be obtained from the rotation curve derived in the Weyl model. MOND on the other hand predicts the BTFR from the theory once the force law is modified from Newton's law in the low acceleration regime,
i.e. $g_{_{bar}} \propto g_{_{obs}}^2$ is assumed as in \ref{Eq2}.

We further note that the theoretical constraints on the value of $\alpha_{DM}$ ($\alpha_{DM}<0$ or $0<\alpha_{DM}<1$) makes it easier to reproduce the Radial Acceleration Relation. Had $\alpha_{DM}$ been greater than unity then we would not obtain $g_{_{obs}} \propto g_{_{bar}}^{1/2} $ in the limit $g_{_{bar}}<< 10^{-10} \rm m~ s^{-2}$. Similarly , the requirement $\beta_{DM}<<1$ allows the acceleration scale $\frac{GAc^4\beta_{_{DM}}^2}{R^2_{c(DM)}}$ obtained from the Weyl model to be of the same order as the MOND universal acceleration scale. Therefore, the theoretical constraints on the Weyl parameters $\alpha_{DM}$ and $\beta_{DM}$ are consistent with the predictions of the RAR.

However, the rotation curve cannot be predicted from the observed mass distribution in the braneworld scenario. The density profile presented in \ref{46} corresponds to the density profile of only the Weyl fluid and carries no information about the observed mass distribution. However, if we assume that the galaxy obeys the BTFR then the flat part of the rotation curve which is related to the parameter $\beta_{DM}$, can be fixed. However, there are other unknown parameters, e.g.  $\alpha_{DM}$ and $R_c$ (\ref{Eq8}) which can be determined only after fitting the full rotation curve. In this context it is important to mention that MOND predicts the rotation curve \cite{Lelli:2017vgz} due to the presence of the universal acceleration scale $g^\dagger$ (\ref{Eq3}).

\section{Can Weyl fluid act as a proxy for dark matter for LSB galaxies?}
The low surface brightness galaxies (LSBs) are dark matter dominated with negligible baryonic mass such that the effect of dark matter continues to dominate even in the 
region where the baryonic disc of stars and gas is present  \citep{Gergely:2011df,
de2001mass, banerjee2010dark, banerjee2017mass}.
In this work however, we do not assume dark matter, but attribute its origin to higher dimensional gravity. Such LSB galaxies have a constant mass density core with core radius of a few kpc \citep{de2005halo}. Unlike High Surface Brightness (HSB) galaxies, for LSBs we ignore baryonic contribution even within the core radius.
Assuming the core radius to be $R_{c(DM)}$ and the mass of the core to be $M_{DM}$, the constant density of the core is then given by $\rho_{c_{DM}}=\frac{3M_{DM}}{4 \pi  R^{3}_{c(DM)}}$.

It therefore follows from \ref{46} that the density profile describing the dark matter in low surface brightness galaxies assumes the form,
\begin{align}
\rho_{_{DM}}(r)&=\frac{3M_{DM}}{4 \pi  R^{3}_{c(DM)}}(1-H_{k_{DM}}(r)) ~+~ \nonumber \\
&H_{k_{DM}}(r)\left \lbrace \frac{c^{2} \beta_{DM}}{Gr^{2}}\left(1-\alpha_{DM} \bigg(\frac{R_{c(DM)}}{r}\bigg)^{1-\alpha_{DM}}\right )\right\rbrace
\label{47}
\end{align}
where, $H_{k_{DM}}$ is a smoothening function given by \citep{Gergely:2011df},
\begin{equation}
H_{k_{DM}}(r)=\frac{1}{1+\rm{exp}(-2k_{DM}(r-R_{c(DM)}))} 
\label{48}
\end{equation}
such that it smoothly approaches the Heaviside step function as $k_{DM}$ tends to infinity \cite{Gergely:2011df}, i.e.,
 \[    H(R_{c(DM)})= \lim_{k_{DM} \to \infty} H_{k_{DM}}(R_{c(DM)})=\left\{
                \begin{array}{ll}
                  0 ~~~~~r<R_{c(DM)}\\
                  1 ~~~~~r\geq R_{c(DM)}\\
                \end{array}
              \right.
  \]
The purpose of introducing the smoothening function is to ensure that the density distribution of the Weyl fluid changes smoothly from the region of constant density core to the constant rotation curve regime, such that there is no abrupt jump in the energy-momentum tensor of the Weyl fluid across the surface of the core (i.e. no distributional source layer on the surface of the sphere of radius $R_{c(DM)}$). This is necessary, because the metric (given by \ref{15}) is the same both inside and outside the core and hence there is no discontinuity in its first derivative. This implies that the extrinsic curvature is continuous across the core and not only the Israel junction conditions but also the Lichnerowicz continuity conditions hold \cite{Gergely:2011df}.

\subsection{2-component model of the baryonic disc}

We model the galactic baryonic disc as composed of two concentric, co-planar, axi-symmetric discs of stars and atomic hydrogen (HI) gas,
which are gravitationally coupled to each other and also in the external force field of a rigid halo of the dark mass. Molecular hydrogen gas H$_2$ 
has been neglected in this dynamical model as it is hardly traced in LSBs. In fact, the H$_2$ to HI mass ratio in late-type LSB spirals is reported to be $\approx 10^{-3}$
\citep{matthews2005detections} and hence can be considered to be
a dynamically insignificant component in UGC7321 without significant errors.

The joint Poisson equation in terms of the galactic cylindrical coordinates $(R,\phi,z)$ is given by
	\begin{equation}
	\frac{\partial ^{2} \Phi _{\rm{total}}}{\partial z^{2}} +\frac{1}{R} \frac{\partial}{\partial R}(\frac{R \partial \Phi _{\rm{total}}}{\partial R}) =
	4 \pi G(\sum_{i=1}^{2}\rho _{i} + \rho_{DM})
	\end{equation}
	Here, $\Phi _{\rm{total}}$ is the total gravitational potential due to the 
	baryonic disc (stars + gas) and the halo of the dark mass. $\rho_{i}$ represents the density of the $i^{th}$ component ($i$ = (stars), HI) and $\rho_{\rm{DM}}$ is the effective density of the dark mass as given by \ref{47}. We note that $\rho_{\rm{DM}}$ is characterized by five free parameters. Here, the angular term drops off due to the assumed azimuthal symmetry and, similarly, the radial term, for a flat rotation curve.
        Thus the joint Poisson's equation reduces to
	\begin{equation}
	\frac{\partial ^{2} \Phi _{\rm{total}}}{\partial z^{2}}  =
	4 \pi G(\sum_{i=1}^{2}\rho _{i} + \rho_{\rm{DM}})
	\end{equation}
	
We further assume that the stars and HI gas are in vertical hydrostatic equilibrium and also their vertical velocity dispersions remain constant in the $z$-direction. At a given galacto-centric radius $R$, the equation of vertical hydrostatic equilibrium for the $i^{\rm{th}}$ component of the disc (i=(stars), gas) is given by  
\begin{equation}
\frac{\sigma_{z,i}^2}{\rho_i}\frac{\partial \rho_i}{\partial z} + \frac{\partial \Phi _{\rm{total}}}{\partial z} =0
\end{equation} 
where $\sigma_{z,i}$ the vertical velocity dispersion of the
$i^{\rm{th}}$ component \citep{rohlfs1977lectures}.	

The radial profile of the vertical velocity dispersion of the stars is modelled as 
	\begin{equation}
	\sigma_{\rm{z,s}}(R)=\sigma_{0\rm{s}}\rm{exp}(-R/\alpha_{\rm{s}}R_D)
	\end{equation} 
where $\sigma_{0\rm{s}}$ and ${\alpha}_{\rm{s}}$ are free parameters. Here $\sigma_{0\rm{s}}$ denotes the vertical velocity dispersion of the stars at the centre of the galactic disc; ${\alpha}_{\rm{s}}$ represents the scale length (in units of the exponential disc scale length $R_D$) with which the vertical velocity dispersion falls off with $R$. This is closely following \cite{van1989photometry} who found that a vertical velocity dispersion of an isothermal, self-gravitating stellar disc should fall off exponentially with radius with a scale length of 2 $R_D$ to match the observed constant stellar scale height in edge-on galaxies.

Similarly, the vertical velocity dispersion of HI has in
general been found to either remain constant, or to linearly
vary with radius. See, for example, \citet{narayan2002vertical}.
Therefore, we parametrize the vertical velocity dispersion of
HI as a polynomial as follows:

\begin{equation}
	 \sigma_{\rm{z,HI}}(R) = \sigma_{0HI}+\alpha_{HI} R +\beta_{HI} R^{2}
\end{equation}
where $\sigma_{0HI}$, $\alpha_{HI}$ and $\beta_{HI}$ are free parameters. Here $\sigma_{0HI}$ is the HI vertical velocity dispersion at the galactic centre.

	Therefore, at a given galacto-centric radius $R$, combining the joint Poisson's equation and the equation for vertical hydrostatic equilibrium for the $i^{\rm{th}}$ component we get
	\begin{equation}
	\frac{\partial^{2}\rho_{i}}{\partial z^{2}} = -4\pi G
	\frac{\rho_{i}}{{\sigma_{z,i}}^2} (\rho_{i} +  \rho_{\rm{DM}}) +
	(\frac{\partial\rho_i}{\partial z})^2 \frac{1}{\rho_{i}};
	\label{54}
	\end{equation}
	 
 Since the stars and gas in the galaxy are mostly concentrated in the galactic disk, it is more natural to express the equations governing their density distributions in cylindrical polar coordinates. However, the vertical density distributions of the stars and gas depend on the density distribution of the dark matter halo $\rho_{DM}$ which has a spherical symmetry (see discussion in \ref{S2-1}). Since the equations governing the density profiles of baryons (\ref{54}) are expressed in cylindrical coordinates but the density profile for the Weyl fluid (\ref{47}) is expressed in spherical polar coordinates, the spherical radius $r$ (in $\rho_{DM}$) is expressed in terms of the cylindrical radius $R$ and the vertical height $z$, in order to solve the two coupled differential equations for stars and gas at every cylindrical radius $R$. 

For a given set of free parameters
($M_{DM}$, $R_{c(DM)}$, $\alpha_{DM}$, $\beta_{DM}$, $k_{DM}$, $\sigma_{0\rm{s}}$, ${\alpha}_{\rm{s}}$, $\sigma_{0HI}$,
$\alpha_{HI}$, $\beta_{HI}$),
the  above set of two coupled, non-linear, ordinary differential equations (\ref{54}) in the variables $\rho_{\rm{s}}$ and $\rho_{\rm{HI}}$ is solved iteratively using Runge-Kutta method with initial conditions at mid plane $z=0$.See, for example, \citet{narayan2002vertical} for details. The half-width-at-half-maxima of the density distribution $\rho_i$ at a given $R$ thus obtained is taken to be the scaleheight $h_z$ at that $R$ for that i$^{\rm{th}}$ component.
The set of parameters which gives the best match to the observed stellar and HI scaleheight versus $R$ data defines our best-fitting model.

We use the \emph{Markov Chain Monte Carlo (MCMC)} method for determining the best-fitting set of parameters of our model. We use the task \emph{modMCMC} from the publicly available $R$ package \emph{FME} \citep{soetaert2010inverse}, which implements MCMC using adaptive Metropolis procedure \citep{haario2006dram}.

\subsection{Input Parameters}

UGC 7321 is a prototypical superthin galaxy with a radial-to-vertical axis ratio of 10.3. It is observed at an inclination of $88^{\circ}$ and is at a distance of 10 Mpc \citep{matthews2000h,matthews1999extraordinary}. The galaxy has a steeply rising rotation curve with an asymptotic velocity of about 110 km s$^{-1}$ \citep{uson2003hi}. 

\begin{table}
\begin{minipage}{100mm}
\large
\hfill{}
\caption{Stellar parameters of UGC 7321 in B-band.}
\begin{center}

\centering
\begin{tabular}{|l|c|}
\hline
Parameters& $UGC7321^{B}$   \\
\hline    
\hline
$\mu_{0} (\rm{mag} arcsec^{-2})$ \footnote{Central surface brightness of stellar disk} & 23.5   \\
$\Sigma_{0} (M_{\odot}{\rm{pc}}^{-2})$  \footnote{Central surface density of the stellar disk}  & 34.7 \\
$R_{D} (\rm{kpc})$ \footnote{Disc scalelength of the exponential stellar disk} & 2.1  \\
$h_{z} (\rm{kpc})$    \footnote{Scaleheight (HWHM) of the stellar disk}  &  0.105 \\
$M_{stars}$\footnote{Stellar mass calculated using $2\pi\Sigma_{0}R^{2}_{d}$} &$9.6 \times 10^{8} M_\odot$ \\
$M_{HI}$\footnote{HI mass \cite{uson2003hi}}& $1.1\times10^{9} M_\odot$\\
\hline
\end{tabular}
\hfill{}
\end{center}
\end{minipage}
\end{table}
The de-projected central surface brightness in B-band is $\mu_0  \approx 23.5\rm{mag}\rm{arcsec}^{-2}$ \citep{matthews1999extraordinary}. It also has high values of the dynamical mass-to-light ratio
with $M_{\rm{dyn}}/M_{\rm{HI}}=31$ and $M_{\rm{dyn}}/M_{L_{B}}=29$, where $M_{L_{B}}$ is the B-band luminosity, and $M_{HI}$ the total HI mass is the galaxy, $M_{\rm{dyn}}$ being the dynamical mass of the galaxy. This indicates that the galaxy is dark matter rich as was also corroborated by \cite{banerjee2013some}, who found dark matter dominates the disc dynamics at all radii in this galaxy. Earlier studies have shown that in the optical i.e., B-band , the surface density profile is well-fitted with an exponential. The same trend holds good for UGC7321 and hence  ${\Sigma}_i (\rm{R})$ is given by 
$${\Sigma}_s (R) = {\Sigma}_0 \rm {exp} (-R/R_D) $$ 
where ${\Sigma}_s (0)$ is the central stellar surface density and $R_D$ the exponential stellar disc scalelength. The structural parameters corresponding to the 
exponential stellar disc of UGC 7321 in B-band were either directly taken or derived from \cite{uson2003hi}. 

\begin{table}
\begin{minipage}{100mm}
\large
\centering
\caption{Input parameters for HI}
\begin{center}
\begin{tabular}{|l|c|}
\hline
Parameters& UGC7321 \\ 
\hline    
\hline
$\Sigma_{01} (M_{\odot}{\rm{pc}}^{-2})$ \footnote{Central surface density of the first HI gaussian disk}  &  4.912   \\
$\Sigma_{02} (M_{\odot}{\rm{pc}}^{-2})$ \footnote{Central surface density of the second HI gaussian disk}   & 2.50       \\
$a_{1}$ ({\rm{kpc}}) \footnote{Centre of the first HI gaussian disk}  & 3.85\\
$a_{2}$ ({\rm{kpc}}) \footnote{Centre of the second HI gaussian disk} & 0.485  \\
$r_{01}$ ({\rm{kpc}})\footnote{Scalelength of the first HI gaussian disk } &2.85  \\
$r_{02}$ ({\rm{kpc}})  \footnote{Scalelength of the second HI gaussian disk}  &1.51 \\
\hline
\end{tabular}
\hfill{}
\end{center}
\end{minipage}
\end{table}

The  HI surface density and the HI scaleheight data for UGC 7321 were obtained from \cite{uson2003hi} and \cite{o2010dark} respectively. Earlier work indicated that the radial profiles of HI surface density could be well-fitted with double-gaussians profiles. See, for example \cite{begum2005dwarf}, \cite{patra2014modelling}, possibly signifying the presence of two HI discs. Also, galaxies with the HI surface density peaking away from the centre are common, which indicates the presence of an HI hole at the centre.The HI surface density profiles could be fitted well with off-centred double Gaussians given by
$$ {\Sigma}_{HI} (R) = {\Sigma}_{01} \rm{exp} \Big[ -{\frac{{(r-a_1)}^2}{2 {r_{01}}^2}}\Big] + {\Sigma}_{02} \rm{exp}\Big[-{\frac{{(r-a_2)}^2}{2 {r_{02}}^2}}\Big] $$
where ${\Sigma}_{01}$ is the central surface density, $a_1$ the centre and $r_{01}$ the scalelength of gas disc 1 and so on.

The parameters corresponding to stellar disc and the HI disc are summarized in Tables 1 and 2 respectively.



\section{Results}

\begin{figure*}
\begin{center}
\begin{tabular}{ccc}
\resizebox{57mm}{50mm}{\includegraphics{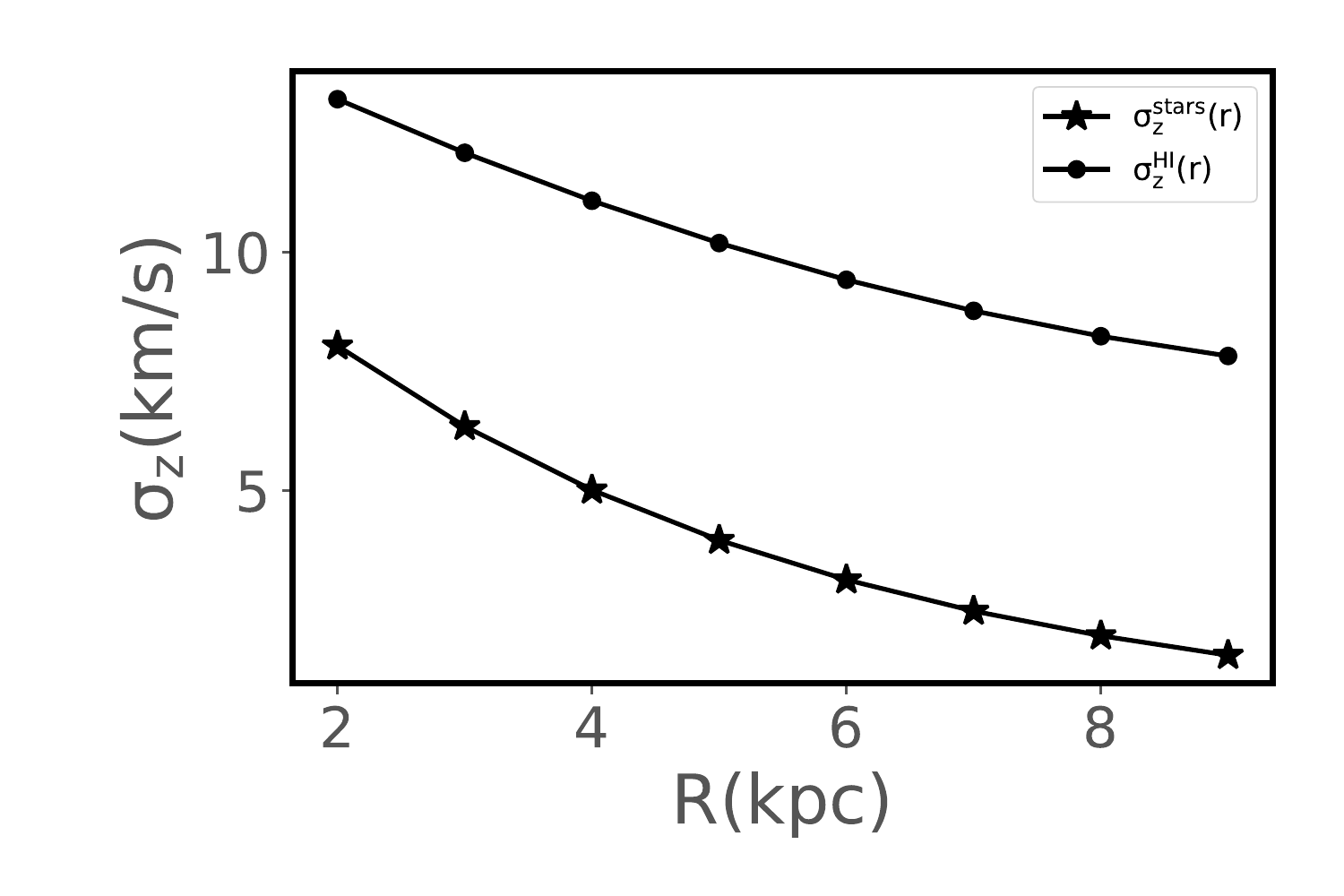}} &
\resizebox{57mm}{50mm}{\includegraphics{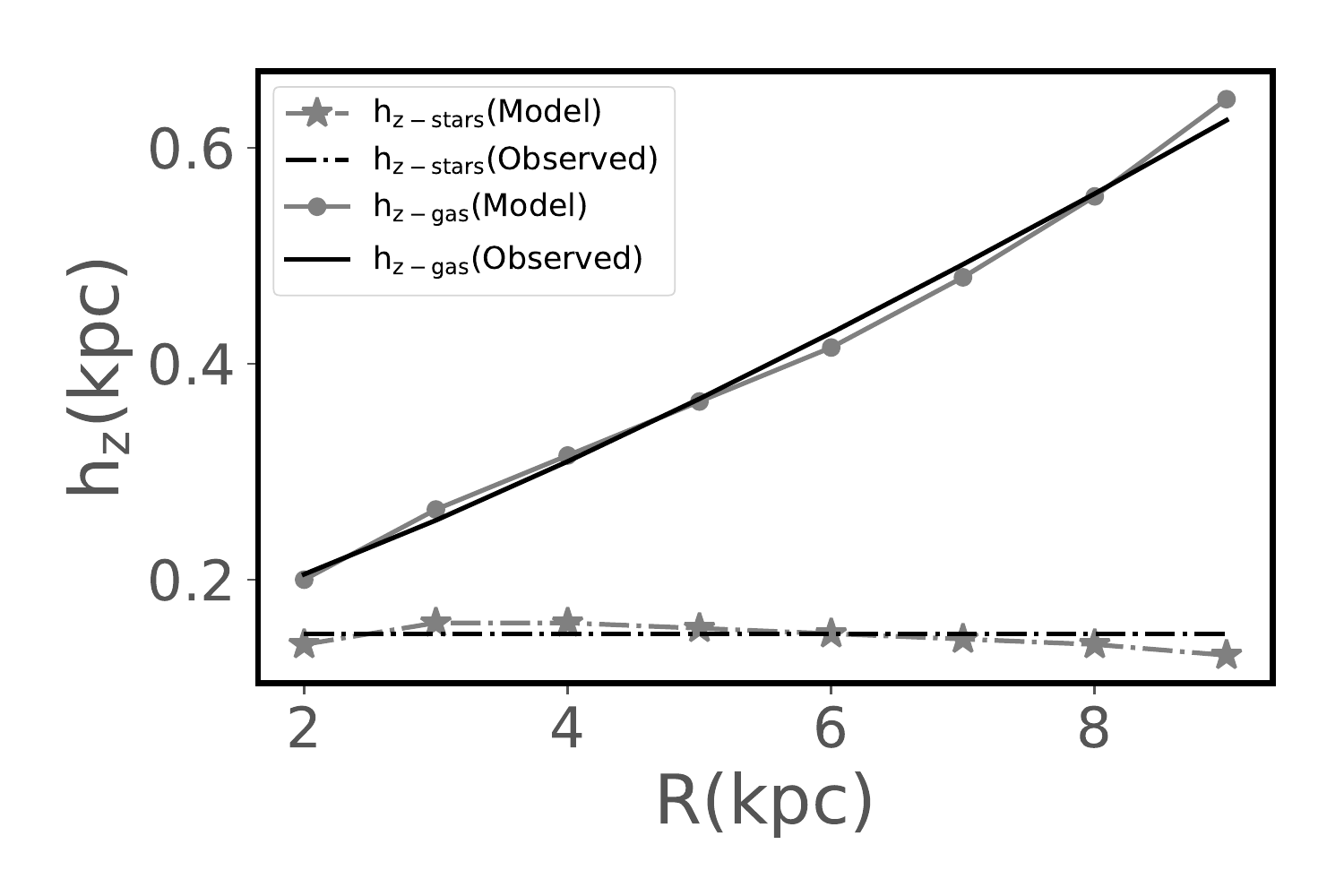}} &
\resizebox{57mm}{52mm}{\includegraphics{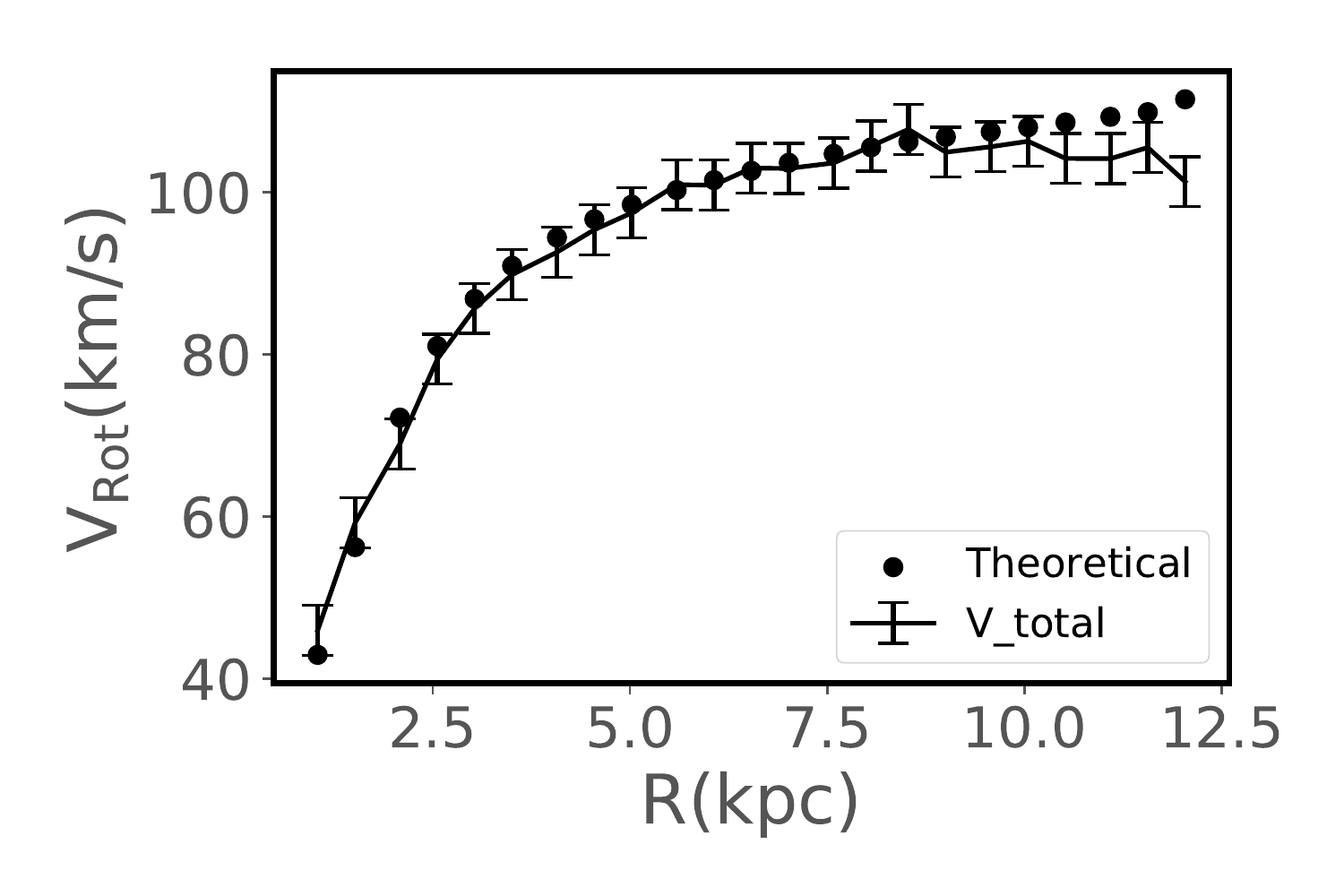}} \\
\end{tabular}
\end{center}
\caption{In the Left Panel, we present the vertical velocity dispersion of stars and
HI as a function of galacto-centric radius $R$ using `stars' and `dots' respectively, as determined from the 2-component model of the baryonic disc embedded in a halo of dark mass as constrained by the observed stellar and HI scaleheight data. In the Middle Panel, we plot the observed stellar and HI scaleheight with black `dash-dot' and black `solid line' respectively. Overlaid on them is the modelled stellar and HI scaleheight represented by grey `stars' and `dots' respectively.
In the Right Panel, we present the rotation curve of UGC 7321 using the best-fit parameters describing the dark matter density profile in the brane-world model and superpose it on the observed rotation curve.}
\end{figure*}

Our dynamical model of UGC 7321 in $B$-band consists of ten free parameters. Out of these, five free parameters correspond to the baryonic disc: $\sigma_{0\rm{s}}$ and ${\alpha}_{\rm{s}}$ corresponding to the stellar vertical velocity dispersion profile and $\sigma_{HI}$,  $\alpha_{HI}$ and $\beta_{HI}$ corresponding to the HI vertical velocity dispersion profile. The remaining five free parameters are related to the dark mass profile (\ref{47}) derived from the brane world model: $M_{DM}$, $R_{c(DM)}$, $\alpha_{DM}$, $\beta_{DM}$ uniquely describing the Weyl fluid while $k_{DM}$ is associated with the smoothing function. We use the observed stellar and HI scaleheight to constrain the ten free parameters described above. 

In Figure 1 [Left Panel], we present the vertical velocity dispersion of stars and
HI as a function of galacto-centric radius from the 2-component model. We find that the central stellar velocity dispersion is (13.4$\pm$0.6) kms$^{-1}$, which falls off exponentially with a scalelength (2.1$\pm$0.4)$R_{D}$, $R_{D}$ being the exponential stellar disc scale length. The central HI vertical velocity dispersion is given by (15.4$\pm$0.5) kms$^{-1}$ with $\alpha_{HI}=(-1.3\pm 0.2)$ kms$^{-1}$kpc$^{-1}$ and $\beta_{HI}=0.04\pm 0.02$ kms$^{-1}$kpc$^{-2}$, indicating that it falls off almost linearly with radius. We note that the velocity dispersion of the stars is lower than the HI velocity dispersion. It is, in general, not possible for the stars to have lower dispersion than the gas clouds in which they are formed, as stars are collision-less and hence cannot dissipate energy via collisions. This possibly indicates that the thin disc stars were born in an underlying cold component of the gas with lower values of velocity dispersion \citep{patra2014modelling}.

\begin{figure*}
\vspace{-1.5cm}
\hspace*{-1.6cm}
\resizebox{160mm}{145mm}{\includegraphics{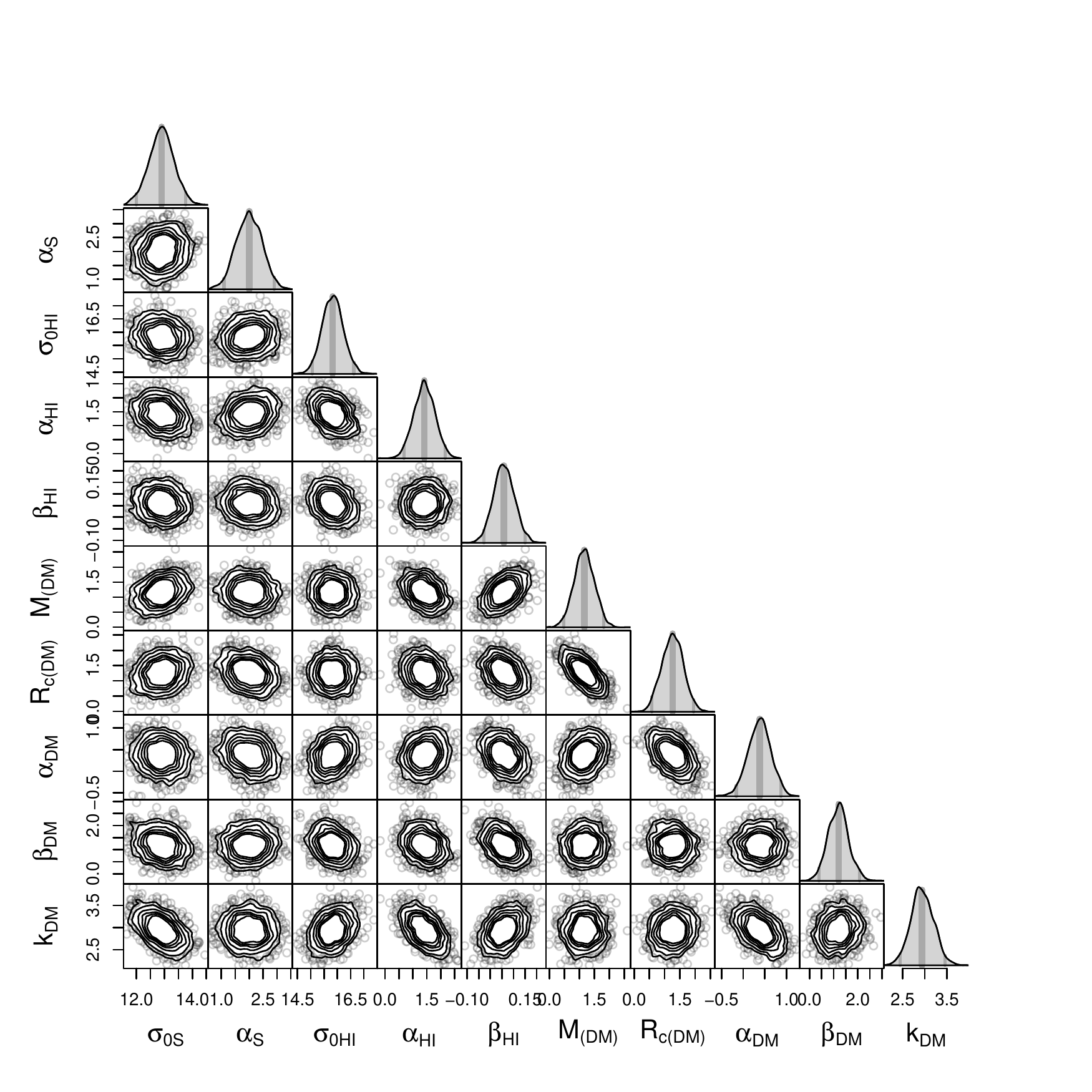}}
\caption{We present the correlation plots and the posterior distributions of the free parameters characterizing the dynamical model of UGC7321 consisting of the 2-component baryonic disc embedded in a halo of dark mass as constrained by the observed stellar and HI scaleheight data using the MCMC method.}
\end{figure*}

The values of dark matter mass obtained from the model $M_{DM}=(1.3\pm 0.2) \times 10^{9} M_{\odot}$, core radius $R_{c(DM)}=(1.3\pm 0.2)kpc$, the parameters describing the Weyl fluid $\alpha_{DM}=-0.4\pm 0.1$, and $\beta_{DM}=(1.4 \pm 0.2) \times 10^{-7}$, the term associated with smoothing function is given by $k_{DM}=2.7\pm0.3$. The results are summarized in Table 3. We note the the estimated dark mass $M_{DM}$ is consistent with the typical mass of the dark matter in low surface brightness superthin galaxies. The value of $M_{DM}$ is of the order of $10^{9}$, which is in agreement with the order of magnitude of dark masses of LSBs reported in \citet{gergely2011galactic}. The core radius $R_{c(DM)}$ is 1.3 kpc ($\sim$0.6 $R_D$), possibly indicating that the Weyl fluid has a dense and compact structure. This complies with a previous study by \citet{banerjee2010dark} showing UGC7321 has dense and compact pseudo-isothermal dark matter halo. The values of $\alpha_{DM}$ and $\beta_{DM}$, which refer to the deviations of the spherically symmetric
metric from the Schwarzschild scenario in general relativity, satisfy the condition $\alpha_{DM}<0$ and $0<\beta_{DM}<1$ and are also comparable with the values of the parameter set obtained by \cite{gergely2011galactic} for a sample of nine LSBs using the observed rotation curve. 
However, we note that the value of $k_{DM}$ is of the order of 15-150 $kpc^{-1}$, for the sample LSBs studied by \cite{gergely2011galactic}, but for UGC 7321 we find that  the value of $k_{DM}$ equals $2.66 kpc^{-1}$. This could be possibly attributed to the fact that UGC7321, like some other superthins, has a relatively steeply rising rotation curve compared to  LSBs in general \citep{banerjee2016mass}.

In Figure 1 [Middle Panel] we have overlaid the scale heights predicted  by the 2-component 
model on the observed scale heights. This confirms that our best-fitting model matches well with the observations. In Figure 1 [Right Panel] we compare the theoretical rotation curve corresponding to the best-fitting parameters of the 2-component model with the observed rotation curve of UGC 7321. The figure illustrates that the theoretical rotation curve obtained by using the best-fitting parameters from the scaleheight constraint on two-component model mostly agrees with the observed rotation curve within error-bars. We note that the rotation curve begins to flatten from $\sim 10$ kpc and continues to be flat as far as we have available data. 

However, we expect galactic dark matter halos to have a finite "size", which should result in a falling rotation curve beyond the radius equal to the ``size" of the galactic halo. One way to indicate the "size" of the dark matter halo is to consider its virial radius, which is the radius at which the average dark matter density is 200 times the critical density of the universe. For UGC 7321, this is $\sim$70 kpc \citep{de2001mass, oh2008high, lelli2016sparc}. However, with the current observations, there is no way to check if the rotation curve actually falls off beyond the virial radius of the galaxy due to the near-absence of tracer gas at those large radii.

Since we do not have data points of the rotation curve at large galacto-centric radii, the brane-world model parameters are constrained by the observed rotation curve only, which extends upto a radius of about one-tenth of the virial radius of the galaxy. If we extend the rotation curve of UGC 7321 using the best-fitting values of the model parameters (given in Table 3 of the revised version) say upto, $\sim 100$ kpc, then we note that the curve continues to be flat with no decline.  However, this asymptotic behavior is consistent with the pseudo-isothermal dark matter density profile which is routinely used to fit the rotation curves of all galaxies \citep{de2001mass, oh2008high, lelli2016sparc}.

In Figure (2), we present the correlation plots and the posterior distributions obtained from the MCMC fitting of the 2-component model. Except for a few cases, we do not note strong correlations between the free parameters of the model.

\begin{table}
\begin{minipage}{100mm}
{\small
\hfill{}}
\centering
\caption{Best-fit parameters found after optimizing the model}
\begin{center}
\begin{tabular}{|l|c|}
\hline
Parameters & UGC7321 \\ 
\hline    
\hline

$\sigma_{0s} $ (\rm{kms}$^{-1}$)\footnote{Central stellar vertical velocity dispersion}  & $13.4 \pm 0.6$      \\
$\alpha_{s} $ $(\rm{kpc}^{-1})$ \footnote{Exponential radial scale length (in units of R$_D$) of the stellar vertical velocity dispersion}   & $2.0 \pm 0.4$        \\
$\sigma_{HI}$ ($\rm{kms}^{-1}$) 
\footnote{Central HI vertical velocity dispersion}  & $15.4 \pm 0.5$  \\
$\alpha_{HI}$ $(\rm{kms}^{-1}\rm{kpc}^{-1})$ \footnote{Radial gradient of HI vertical velocity dispersion} &  $-1.3 \pm 0.2$ \\
$\beta_{HI}$ $(\rm{kms}^{-1}\rm{kpc}^{-2})$ \footnote{Gradient of the radial gradient of HI vertical velocity dispersion} & $0.04 \pm 0.02$ \\
$M_{DM}$ ($M_{\odot}$)  \footnote{Dark mass}  & $(1.3 \pm 0.2) \times10^{9}$\\
$R_{c(DM)}$ (\rm{kpc})  \footnote{Core-radius of the dark mass density profile}  &  $1.3 \pm 0.2$\\
$\alpha_{DM}$  \footnote{Weyl-fluid parameter}  & $-0.4 \pm 0.1$ \\
$\beta_{DM}$   \footnote{Weyl- fluid parameter}  & $(1.4 \pm 0.2) \times 10^{-7}$\\
$k_{DM}$ $(\rm{kpc}^{-1})$  \footnote{Smoothing term associated with dark mass density profile}  & $2.7 \pm 0.3$\\
\hline
\end{tabular}

\end{center}

\end{minipage}
\end{table}

\section{Summary and Conclusions}
In this work, we explore the possibility of higher dimensional gravity in explaining the vertical scaleheight structure of the dark matter dominated LSB galaxy UGC 7321, where the role of dark matter is played by five dimensional Einstein gravity.  
The five dimensional Einstein's equations are projected onto the 3-brane where our visible universe resides such that the effective 4-dimensional gravitational field equations inherits a source term originating from the bulk.
The source term owes its origin to the electric part of the bulk Weyl tensor and captures the non-local effects of the bulk onto the brane. For a brane observer therefore, the bulk effectively behaves like a fluid (the so called Weyl fluid) possessing an energy density and pressure, viz, the dark radiation and the dark pressure terms respectively.
Due to the presence of the Weyl term, the static, spherically symmetric and asymptotically flat solution of the 4-d effective gravitational field equations deviates from the Schwarzschild spacetime. The degree of deviation from the Schwarzschild scenario is determined by the parameters connecting the dark radiation and the dark pressure terms in the equation of state of the Weyl fluid. 
The equation of state essentially assigns some initial conditions to the evolution equations of the off-brane quantities (e.g. the normalized normals) along the extra dimensions. 
In the context of galaxies, it turns out that the appropriate equation of state is $p=(a-2)\mu-B$, where $a=3$ and $B=0$ corresponds to the Schwarzschild scenario.      
With this equation of state and by employing $\tilde{q}\equiv GM/R \approx 10^{-7}<<1$ (which holds true in the galactic situation), one can obtain the density 
profile and the rotation curve of the LSB galaxies in terms of the equation of state parameters. These are fitted with the available rotation curve and 
scale height data of UGC 7321 to constrain the Weyl parameters as well as the stellar and HI vertical velocity dispersion profiles. 
It is important to note that the Weyl model cannot predict the rotation curve from the observed mass distribution since the Weyl parameters are not
universal constants. In other words, by fitting the rotation curves and the vertical scale height data of different galaxies we obtain different sets of values for the Weyl parameters within their physically allowed domain.

The rotation curve can be predicted in the premise of MOND due to the universality of the acceleration scale $g^\dagger$. However, unlike the 
braneworld model, the physical origin of the universal acceleration scale, the interpolating function between the low and high accelerations or the force law
obeyed by particles in the deep-MOND regime is not well understood \cite{Milgrom:2014usa}. As far as compliance of MOND with the 
vertical structure of galaxies is concerned, we may note that \cite{sanchez2008thickness} studied if the observed HI vertical 
thickness of the Milky Way could be modelled in the MOND scenario. Using a 2-component galactic disc model of gravitationally-coupled stars and 
gas, similar to the one studied in this paper, and assuming a constant value of 9 kms$^{-1}$ for HI vertical velocity dispersion, 
they found that the model scaleheight fitted the observed scaleheight well at $R > 17$ kpc i.e., beyond 5-6 $R_D$. However, at $R < 17$ kpc, 
the model was found to under-predict the observed scaleheight by about 40$\%$.

By fitting the rotation curves of LSB as well as HSB galaxies \citep{Chakraborty:2015zxc,gergely2011galactic} the Weyl parameters have 
been determined in the past.
This motivates us to
confront this model with the stellar and HI vertical scale height data of dark matter dominated LSB galaxies, e.g. UGC 7321   
which is a very well studied object with a dynamical mass as large as $M_{dyn}/M_{HI}=31$ and $M_{dyn}/M_{L_{B}}=29$. Our analysis reveals 
that the Weyl model can not only address the observations related to the rotation curve but can also successfully explain the vertical scaleheight 
data of UGC 7321 within the error bars. Therefore apart from the rotation curve, this work opens up a new observational avenue which can be 
utilized in understanding the role of extra dimensions or other alternative gravity models in the galactic scale. 

\section{Data Availibity}
The data underlying this article are available in the article itself

\section*{Acknowledgement}
KA would like to thank Prof. T. Harko for useful discussion.
IB thanks Sumanta Chakraborty for useful discussions at various stages of this work.
AB would like to acknowledge the research grant from DST-INSPIRE Faculty Fellowship (DST/INSPIRE/04/2014/015709) for partially supporting this work.
The research of SSG is partially supported by the Science and Engineering Research Board-Extra Mural Research Grant No. (EMR/2017/001372), Government of India.
The authors would also like to thank the anonymous referee for insightful comments which have helped improve the quality of the paper.
\small{\bibliographystyle{mnras}}
\bibliography{Brane,superthin_check}


\end{document}